\documentclass[sigconf]{acmart}

\AtBeginDocument{%
  }

\setcopyright{acmcopyright}
\copyrightyear{2024}
\acmYear{2024}
\acmDOI{10.1145/3637528.3671596}

\acmConference[KDD '24]{SIGKDD Conference on Knowledge Discovery and Data Mining}{August 25--29,
  2024}{Barcelona, Spain}

\copyrightyear{2024}
\acmYear{2024}
\setcopyright{acmlicensed}
\acmConference[KDD '24] {Proceedings of the 30th ACM SIGKDD Conference on Knowledge Discovery and Data Mining }{August 25--29, 2024}{Barcelona, Spain.}
\acmBooktitle{Proceedings of the 30th ACM SIGKDD Conference on Knowledge Discovery and Data Mining (KDD '24), August 25--29, 2024, Barcelona, Spain}
\acmISBN{979-8-4007-0490-1/24/08}



\copyrightyear{2024}
\acmYear{2024}
\setcopyright{rightsretained}
\acmConference[KDD '24]{Proceedings of the 30th ACM SIGKDD Conference on Knowledge Discovery and Data Mining}{August 25--29, 2024}{Barcelona, Spain}
\acmBooktitle{Proceedings of the 30th ACM SIGKDD Conference on Knowledge Discovery and Data Mining (KDD '24), August 25--29, 2024, Barcelona, Spain}\acmDOI{10.1145/3637528.3671596}
\acmISBN{979-8-4007-0490-1/24/08}

\usepackage{caption}
\usepackage{algorithm}
\usepackage{algorithmic}
\usepackage{xspace}
\usepackage{enumitem}
\usepackage{mathtools}
\usepackage{xcolor}
\usepackage{multirow}
\usepackage{arydshln}
\usepackage{tikz}

\usepackage{subcaption}
\usepackage{booktabs}
\usepackage{balance}

\definecolor{red}{RGB}{198,50,42}\definecolor{agreen}{RGB}{74, 198, 148}
\definecolor{purple}{RGB}{158, 62, 177}
\definecolor{aqua}{RGB}{87, 180, 181}
\definecolor{orange}{RGB}{255,143,40}
\definecolor{amber}{rgb}{1.0, 0.75, 0.0}
\definecolor{awesome}{rgb}{1.0, 0.13, 0.32}
\definecolor{bronze}{rgb}{0.8, 0.5, 0.2}
\definecolor{indigo}{rgb}{0.0, 0.25, 0.42}
\definecolor{heliotrope}{rgb}{0.87, 0.45, 1.0}
\definecolor{forestgreen}{rgb}{0.13, 0.55, 0.13}
\definecolor{ginger}{rgb}{0.69, 0.4, 0.0}
\definecolor{jade}{rgb}{0.0, 0.66, 0.42}
\definecolor{mediumslateblue}{rgb}{0.48, 0.41, 0.93}
\definecolor{mint}{rgb}{0.24, 0.71, 0.54}
\definecolor{mulberry}{rgb}{0.77, 0.29, 0.55}
\definecolor{linkColor}{RGB}{6,125,233}

\newcommand{\Multidataset}[0]{Nested Measurement Dataset\xspace}
\newcommand{\multidataset}[0]{nested measurement dataset\xspace}
\newcommand{\NestedFusion}[0]{Nested Fusion\xspace}

\begin{document}

\title{Nested Fusion: A Method for Learning High Resolution Latent Structure of Multi-Scale Measurement Data on Mars}




\author{Austin P. Wright}
\affiliation{%
  \institution{Georgia Tech}
  \city{Atlanta}
  \state{GA}
  \country{USA}
}

\author{Scott Davidoff}
\affiliation{%
    \institution{Jet Propulsion Laboratory}
  \department{California Institute of Technology}
  \city{Pasadena}
  \state{CA}
  \country{USA}
}

\author{Duen Horng Chau}
\affiliation{%
  \institution{Georgia Tech}
  \city{Atlanta}
  \state{GA}
  \country{USA}
}

\renewcommand{\shortauthors}{Austin P. Wright, Scott Davidoff, and Duen Horng Chau}

\begin{abstract}

The Mars Perseverance Rover represents a generational change in the scale of measurements that can be taken on Mars, however this increased resolution introduces new challenges for techniques in exploratory data analysis. 
The multiple different instruments on the rover each measures specific properties of interest to scientists, so analyzing how underlying phenomena affect multiple different instruments together is important to understand the full picture. 
However each instrument has a unique resolution, making the mapping between overlapping layers of data non-trivial. In this work, we introduce \textbf{Nested Fusion}, a method to combine arbitrarily layered datasets of different resolutions and produce a latent distribution at the \textbf{highest possible resolution}, encoding complex interrelationships between different measurements and scales. 
Our method is efficient for large datasets, can perform inference even on unseen data, and outperforms existing methods of dimensionality reduction and latent analysis on real-world Mars rover data. 
We have deployed our method Nested Fusion within a Mars science team at NASA Jet Propulsion Laboratory (JPL) and through multiple rounds of participatory design enabled greatly enhanced exploratory analysis workflows for real scientists. 
To ensure the reproducibility of our work we have open sourced our code on GitHub at \url{https://github.com/pixlise/NestedFusion}. 

\end{abstract}

\begin{teaserfigure}
    \centering
    \includegraphics[width=0.95\textwidth]{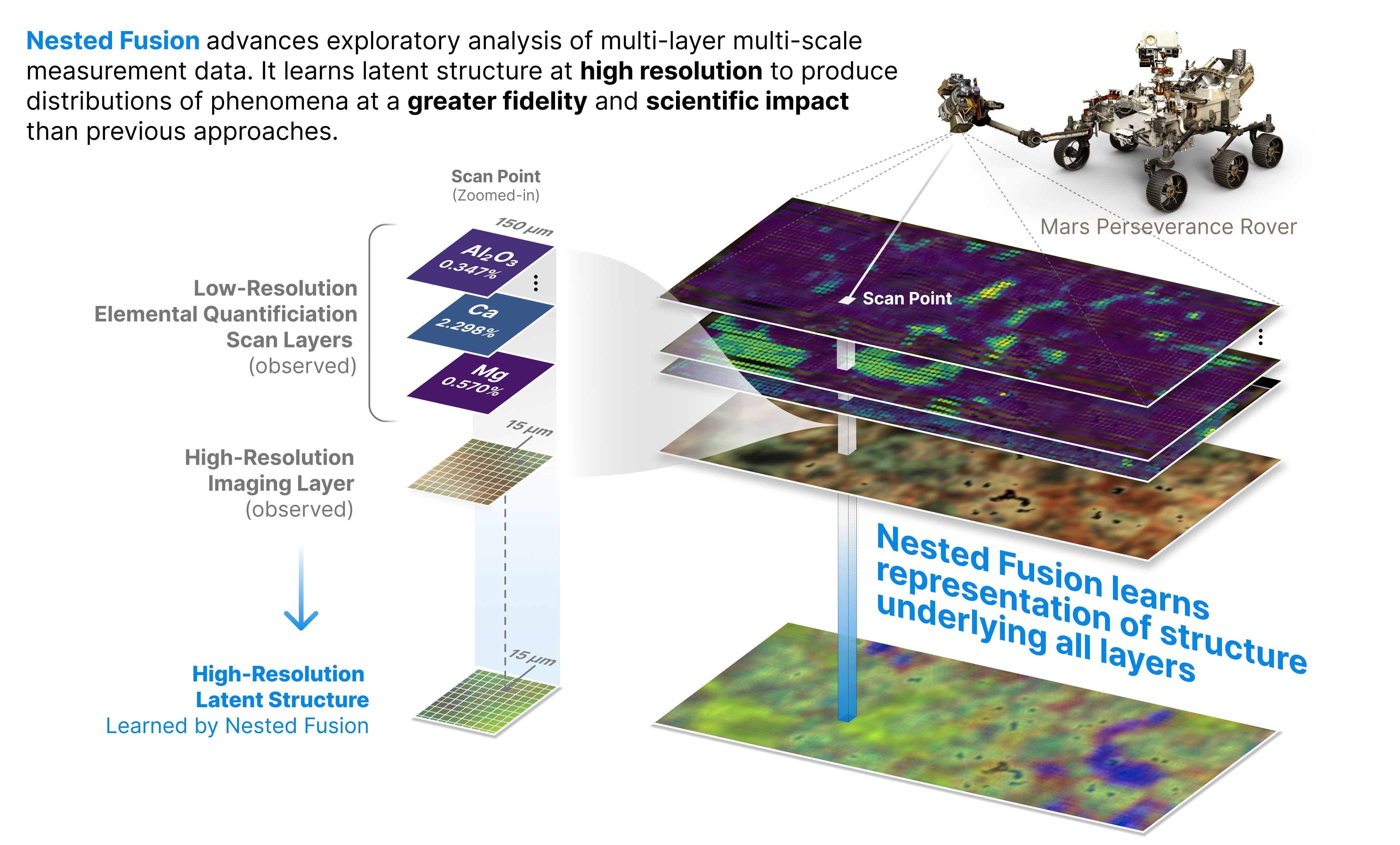}
    \caption{Our method, Nested Fusion, radically accelerates the exploratory analysis of \multidataset{s} by learning the latent structure at high resolution to produce distributions of phenomena at a greater fidelity and scientific impactfulness than previous approaches. In the figure the DOURBES target location is shown, out of over a hundred locations on Mars scanned by the Perseverance Rover at the time of writing.}
    \label{fig:teaser-figure} 
\end{teaserfigure}

\maketitle

\section{Introduction}



In scientific data analysis the initial exploratory phase of visualizing and conceptualizing the relevant empirical phenomena in a dataset is both an essential aspect for effective work and comparatively under studied in the context of scientific applications, where skipping such inductive explorations in favor of immediately utilizing known models for analysis is the de facto standard. 
However, recent work has shown how unanticipated or anomalous phenomena can often mislead such analysis, motivating a workflow that at least starts with purely empirical exploration of data in the initial phases of work after making measurements in order to have a more informed prior of the distribution of actual phenomena within a dataset before applying the more rigorous scientific models to ensure the chosen models are appropriate \cite{wright2023lessons}. 
While common data-centric techniques of exploratory analysis such as dimensionality reduction visualization have proven to be very effective in many domains of scientific inquiry \cite{becht2019dimensionality,bagger2019bloodspot,oetjen2018human,li2019manifold,lin2018understanding,thompson2015automating,pletl2023spectral,gayoso2023deep, weinstein2021structured, xu2021probabilistic,singh2022planetary}, in domains with multiple measurement apparatuses of different resolutions and scales, existing techniques can fail to model some of the phenomena we wish to discover.
This is because the standard formalization for dimensionality reduction techniques is that of a single dataset of measurements of identical shape which corresponds one to one with the set of objects and patterns between objects that the analysis aims to visualize. 
However it is often the case that underlying phenomena are differentiated at levels that do not align with the resolutions of measurement each apparatus perfectly \cite{wright2023lessons}. 
Rather, there may be multiple methods of measurement which each elucidate different aspects of an underlying structure but which all have varying resolution scales and thus are sensitive to the different properties of various aggregations of the structure.  

One such domain where scientists require more powerful exploratory analysis tools is the work done by the PIXL Science team with the Mars Perseverance Rover at NASA (National Aeronautics and Space Administration). 
In service of the high-level goal of searching for signs of a history of life on Mars, scientists are interested in the fine-grain mineral structure of \textit{target} locations on the Martian surface \cite{farley2020}. 
The Perseverance Rover contains two (among many) scientific instruments to assist in this task: the Planetary Instrument for X-ray Lithochemistry (PIXL) instrument \cite{allwood2020}, which includes an X-ray fluorescence (XRF) spectrometer, and a Micro-Context Camera (MCC) for multi-spectral imaging. 
When observing a specific target location of geological interest, the rover will use both of these instruments to conduct two co-aligned scans as shown in Figure \ref{fig:teaser-figure}. 
While both of the instruments scan over the same physical location, their resolutions are much different, where for each scan point, a single XRF spectrum corresponds to a larger patch of approximately 100 MCC imaging pixels. 
At the same time, each instrument elucidates different aspects of the underlying mineralogy of the target. 
While the spatial precision of each MCC pixel corresponds much more closely to individual homogeneous mineral grains, it lacks a nuanced depth of information to accurately differentiate minerals based on chemistry.
On the other hand, each XRF spectrum produces a detailed quantified distribution of the chemical composition of the scan point, but the larger diameter of this point may encompass multiple grains of different minerals thus producing an aggregate chemical distribution. 
The ultimate scientific question is about understanding the distribution of underlying minerals. While both  measurements offer extremely powerful signals concerning this distribution, neither alone encompasses all the possible information to explore, leading to the need for modeling these different measurement scales together.
To tackle these significant scientific challenges, we present the following major contributions:

\begin{enumerate}[topsep=5pt, itemsep=0mm, parsep=2mm, leftmargin=15pt]

\item \textbf{A novel problem formulation} tailored to exploratory analysis  of \multidataset{s}, which consist of irregularly overlapping measurements of multiple scales (Sec \ref{sec:formulation}). 
This formulation is rooted in addressing the practical needs of PIXL scientists at NASA who analyze XRF and MCC data collected by the Mars Perseverance Rover.

\item \textbf{The Nested Fusion algorithm},  a new model for latent analysis and dimensionality reduction for  \multidataset{s} (Sec \ref{sec:algorithm}),
This method is significantly more effective than alternatives, 
yielding latent encodings at a resolution far higher than what existing dimensionality reduction techniques can achieve. 
We evaluate the effectiveness of \NestedFusion both qualitatively within the context of initial data exploration and quantitatively in data reconstruction fidelity. 
\NestedFusion outperforms the state of the art in dimensionality reduction for \multidataset{s}, providing more interpretable and practically useful results (Sec \ref{sec:evaluation}).

\item \textbf{Deployment of Nested Fusion} in scientific practice within the PIXL team for the Mars Perseverance Rover, enabling 
scientifically meaningful visual interpretation
and efficient discovery of cross-modal patterns (Sec \ref{sec:impact}). We analyze how \NestedFusion is utilized in practice and how it fits within the scientists' existing analytic workflows. To ensure reproducibility of our technique and findings, we have open-sourced it at \url{https://github.com/pixlise/NestedFusion}

\end{enumerate}
\section{Background and Related Work}
In this section, we introduce the scientific problem statement and dataset overview from the PIXL instrument on the Mars Perseverance Rover, define our formalization of \multidataset{s}, and go over related work in scientific exploratory data analysis and dimensionality reduction.


\subsection{Mars Perseverance PIXL Data}
\label{sec:dataset-overview}
The PIXL instrument aims to measure the mineral structure of small rock samples (called \textit{targets}) on the surface of Mars contributing toward the larger inquiry towards any potential evidence of a history of life on Mars. 
For each individual target on the martian surface multiple scans are taken. First is the MCC Multi-spectral imaging camera, which takes a series of four images illuminated by specific wavelengths of near-visible light: Near-Infrared (NIR), Green, Blue, and Ultraviolet (UV). This produces a single color image for each target with 4 primary channels, as opposed to the standard 3 channel RGB, and is often analysed using the 16 distinct ratios between them. Each image will contain on average about 500,000 of these 16 channel pixels, spanning a region of approximately 100 square centimeters with each pixel corresponding to a resolution of approximately 15 microns. 
At roughly the same time a scan is taken of the same target with the PIXL instrument for X-Ray spectroscopy. This instrument produces much more detailed quantitative data, consisting of a grid of X-Ray fluorescence spectra which are quantified to represent the distribution of elemental weight percentages at each scan point, we call this distribution a \textit{quantification}. Each scan can consist of between 1000 and 10,000 individual spectra (depending on the particular shape of the target) covering a smaller region of approximately 30 square millimeters. Each \textit{scan point} is measured with a beam diameter of 50-200 microns\footnote{This beam diameter is energy dependent and since there is a nonlinear transformation between the energy levels of the spectrum and the final quantified elemental distribution where each element is quantified using the full energy range, we treat the upper range of the beam diameters as representing the region encompassed in a quantified scan point.}, thus corresponding to a region covering approximately 100 MCC pixels as shown in Figure \ref{fig:teaser-figure}. 

Thus far, at the time of writing, during the time that the Perseverance Rover has been in operation, there have been 103 target locations scanned producing a total of 295,602 52-dimensional (the number of unique elements included in all quantifications) quantified spectra, as well as 26,966,169 MCC pixels. 
However, not all scans include both data types and so for this work focusing on combining information from both measurements, we are restricting to a total of 103,005 scan points which each contain a single quantification as well as 100 corresponding MCC pixels.

\subsection{Related Work}
Previous work in collaboration with PIXL scientists has shown how data science techniques can form an essential component of their scientific workflow by focusing specifically on modeling anomalies and visualizing distinct empirical phenomena\cite{wright2023lessons}. This work focuses on the problem of initial visualization and thus on dimensionality reduction as an effective technique for enabling such visualization for the high dimensional PIXL data.

Dimensionality reduction techniques such as UMAP \cite{mcinnes2020umap}, T-SNE \cite{van2008visualizing}, MDS \cite{kruskal1964multidimensional}, Isomap \cite{tenenbaum2000global}, and the most commonly used PCA \cite{hotelling1933analysis} are fairly ubiquitous in a variety of scientific domains \cite{becht2019dimensionality,bagger2019bloodspot,oetjen2018human,li2019manifold,lin2018understanding}, and even specifically XRF spectroscopy \cite{thompson2015automating}, as well as Mars multi-spectral imaging\cite{pletl2023spectral}. 

Another conceptualization that can produce comparable visualizations is the approach of latent analysis which takes a more, generally Bayesian, probabilistic framework to the problem of learning low dimensional representations. 
These approaches mostly stem from the development of variational autoencoders (VAE) \cite{kingma2013autoencoding}, and different latent models have been introduced to handle many scientific problems \cite{gayoso2023deep, weinstein2021structured, xu2021probabilistic} including planetary science\cite{singh2022planetary} among many other domains.

\begin{table*}
    \centering
    \renewcommand{\arraystretch}{1.5}
    \begin{tabular}{p{0.25\linewidth}  p{0.7\linewidth}}
\toprule
         Symbol / Term & Meaning \\ \midrule
         \Multidataset ($M$) & A class of dataset which combines multiple kinds measurements that cover a common area\\
         Data Scale ($X$) & The set of measurements of a particular kind that define a data layer of a specific resolution\\
         Nested Scale ($X_S$)& A scale at a higher resolution which has a correspondence where multiple measurements in the nested scale correspond to a single measurement in the lower resolution scale\\
         Nesting Function ($\eta$) & A function which maps a specific data point at a scale to the set of data points in the corresponding nested scale that cover the same physical space.\\
         Maximum Resolution Latent Scale ($X_\varnothing$)& The scale for which no further nested scales exist, defining the highest resolution available in the dataset and thus the resolution at which latent structure can be modeled\\
         Latent Base Scale Correspondence ($\beta$) & A function which maps a specific data point at any scale to the set of data points at the maximum resolution latent scale that cover the same space as defined by repeated nesting.\\
         $x_{i} \in X$ & A specific data point at some data scale $X$ \\
         $x^{\varnothing}_{i} \in X_{\varnothing}$ & A specific data point at the maximum resolution latent scale \\
         $z_i$ & The latent encoding corresponding to $x^{\varnothing}_{i}$\\
         \bottomrule
    \end{tabular}
    \caption{Notations and terminology used in this paper}
    \label{tab:notation}
\end{table*}

\section{Proposed Method: Nested Fusion}
\label{sec:method}

Grounded in understanding from previous work with PIXL scientists \cite{wright2023lessons} our aim is to develop a method for visualizing and determining the distribution of mineral phenomena within each PIXL target, and to assist in their identification based on their relationship between the past history of targets. Focusing on targets where both XRF and MCC data are present and overlapping, we hope to enable work to discover new patterns that each individual instrument cannot differentiate independently. 
While scientific interpretation is the end goal, the specific interpretations (i.e., ``we see a grain of olivine here or a potential aqueous intrusion there'') enabled by the method are out of the scope of this work. 
Therefore we introduce a precise formalized problem statement which aims to properly encode the scientific priors and goals of the problem with specific consideration to the non-standard mixed scale measurements present in PIXL data, while simultaneously laying the foundation for how such methods can be more easily generalized to new domains. 
Finally after introducing the problem formulation we will describe our proposed method, \NestedFusion, which looks to solve this problem.

\subsection{Problem Formalization of Nested Measurements}
\label{sec:formulation}
As Figure \ref{fig:teaser-figure} shows the nested hierarchical structure of PIXL data is not immediately amenable to standard data science techniques barring some flattening operation which leads to over aggregation and loss of resolution (see \textit{Joint Models} in Section \ref{sec:alternatives}). Thus we introduce a formalization of \textbf{\multidataset{s}} which we will use to model this structure and subsequently perform better analysis on the data in a more natural manner, while also outlining precisely the requirements that any other dataset must meet in order to utilize the methods introduced in Section \ref{sec:method} in other domains. 
Table \ref{tab:notation} summarizes the notations and terminology introduced in this section and used throughout this paper. 

We recursively define a \multidataset $M$ as consisting of a tuple of two components: $M = (X,S)$. The first component $X = \{ x_{i}\in \mathbf{R}^{d}\}^{N}_{i=1}$ is simply a standard dataset of $N$ independent and identically distributed samples of $d$ dimensional data representing the particular measurements at some specific scale. Then $S$ is what we define as the \textit{nested scale}. The nested scale is a tuple $(M', \eta)$  of another \multidataset $M' = (X',S')$ as well as a \textit{nesting function} ($\eta : X \rightarrow  2^{X'}$) which maps each data point in $X$ to a set of corresponding data points in $X'$ that cover the same underlying physical and latent area. In order to terminate this regress there must be a final scale $X^\varnothing$  which has no further nested scale and thus is notated as $\varnothing$. Having no further nested scale means that $X^\varnothing$ is the highest resolution available in the \multidataset, and so we refer to it as the \textit{maximum resolution latent scale} since our aim to model latent structure at this maximum resolution.

The key assumption is that all of the information at lower resolution scales supervenes on latent information at the maximum resolution. 
That is, that there is some more basic structure underlying the dataset that is approximately modeled at the maximum resolution as an unobserved latent variable, where each sample $x^{\varnothing}_{i} \in X_{\varnothing}$ is generated from a random process involving the latent value $z_{i}$ that has a prior probability distribution $p(z)$, producing some conditional distribution $p(x^{\varnothing}|z)$ that we aim to learn\footnote{Note for notation, we  include indices for actual measurement samples, while not including indices when referring to the random variable that generates the samples.}. For all other scale samples with nesting function $\eta$ we then define the $\beta$ correspondence which returns the set of all latents at the base maximum resolution that correspond to a sample $x_i$ :

\begin{align}
    \beta(x^{\varnothing}_i) &:=\{z_i\}\\
    \beta (x_i) &:= \bigcup_{x'_j\in \eta(x_i)}\beta(x'_j)
\end{align}

\begin{figure*}[]
    \centering
    \includegraphics[width=0.75\textwidth]{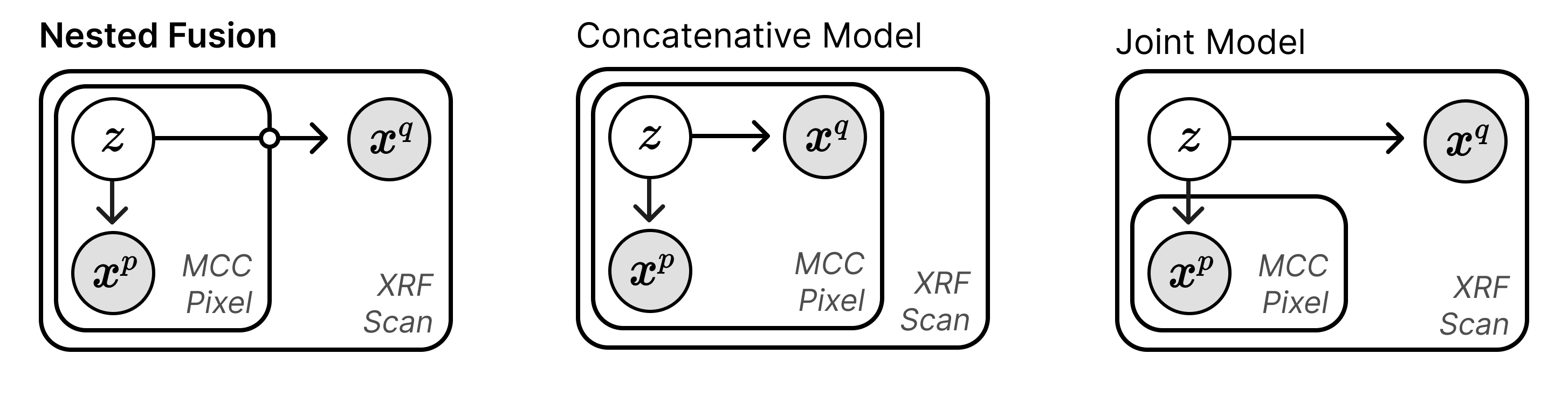}
    \caption{Plate Notation for Graphical Models representing different latent variable formulations for the PIXL MCC \multidataset. From left to right we have: (Left) \NestedFusion, representing the latent corresponding to the maximum resolution datascale and informing higher level measurements through aggregated functions; 
    (Center) the concatenative model where there is a latent at the maximum resolution scale which affects higher level corresponding measurements not in aggregate but independently; and 
    (Right) the joint model where a latent exists at low resolution and determines the whole distribution of all high resolution measurements.}
    \label{fig:pixl-graphical-models}
\end{figure*}

The supervenience assumption then can be restated probabilistically that all lower resolution scale variables are generated from the conditional distributions $p(x|\beta(x))$, and thus are conditionally independent of measurements at any scale other than the maximum. This structure is outlined in the graphical model for the PIXL dataset in Figure \ref{fig:pixl-graphical-models}. 

While seemingly fairly abstract and obscure, this underlying structure and supervenience assumptions of a \multidataset is in fact pervasive in the sciences \cite{sep-reduction-biology,sep-science-mechanisms}. The natural sciences in particular commonly share the physicalist reduction assumption (at least within a single domain), that any given composite object of study is fully reducible to the set of underlying physical objects of which it is composed \cite{sep-physicalism,sep-scientific-reduction}. 
This assumption necessitates that if multiple kinds of measurement apparatus measure an overlapping subject in time and space, then there must be some correspondence relation between the two measurement modalities. Furthermore this assumption enables us to study the intersections between these different layers of composed abstraction, as each class of composite structure is often best observed using separate kinds of measurement that very often do not have perfectly aligned scale and resolution. More complex composite structures will tend to exhibit additional complexity and depth (note the high dimensionality of the PIXL quantified spectra) however at the expense of necessarily being more spatially diffuse. While higher resolution measurements may be possible at the expense of more limited depth.\footnote{Think of the accuracy vs precision distinction, where here generally increasing resolution increases spatial precision but makes it more difficult for each individual measurement to be accurate.}

\subsection{The Nested Fusion Algorithm}
\label{sec:algorithm}

\begin{figure*}[t!]
    \centering
    \includegraphics[width=\textwidth]{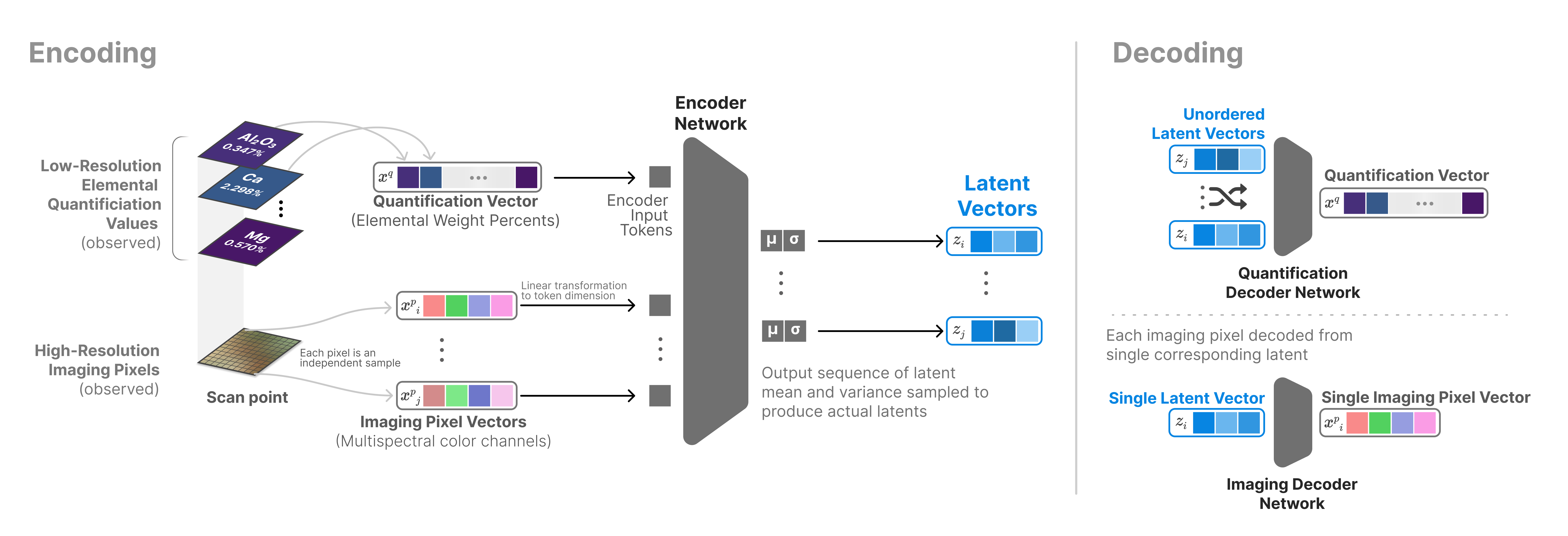}
    \caption{Model architecture and data processing pipeline for \NestedFusion as applied to PIXL data. High resolution latent vectors are encoded given a scan point containing an XRF quantification vector and collection of MCC imaging pixels. 
    }
    \label{fig:model-arch}
\end{figure*}
The previous section describes the formalized problem of learning latent maximum resolution scale variables from nested data. One important aspect to note when introducing our solution is that the formulation of the latent variables at this scale is itself already a modeling approximation. 
In reality we expect fundamental structures within a domain to exist at finer scales than are directly accessible, and so we simply use the highest resolution available in any given \multidataset as a proxy scale for a `true' latent $z$. 
What this lends support to is the use of variational inference as a method to efficiently learn approximate distributions of $z$, which is acceptable as we do not in general actually have strong enough priors about the structure and properties of a `true' $z$ to justify other methods which have significant computational and other drawbacks when compared to widespread empirical success of variational auto-encoding models. 
Therefore the approach taken in this work, \textbf{Nested Fusion}, is a variational auto-encoder model\cite{kingma2013autoencoding} structured to work on \multidataset{s}. 

Figure \ref{fig:model-arch} describes Nested Fusion's architecture. Without loss of generality, we explain how the framework is applied to the PIXL data/scenario presented in Figure~\ref{fig:teaser-figure}. 
Specifically we show how a scan point consisting of both low-resolution elemental quantification values and nested high-resolution imaging pixels, is jointly used to learn high-resolution latent vectors. 
The latents are learned though optimizing via stochastic variational inference\cite{hoffman2013stochastic} both encoder and decoder models to maximally reconstruct the original scan points\footnote{In addition to the other components of the KL Divergence loss used in variational inference which integrates some priors on the latents as well}.
The 1-, 2-, or 3-dimension latents would then be used for visualization by PIXL Scientists.

First, let us consider the encoder step. For the encoder model, 
which estimates the conditional latent distribution given the data $q(z|M)$, 
we must choose a class of distributions for the latent prior $p(z)$ and specify the relevant class of distributions for the data type of each measurement scale. 
We focus on a basic prior model of latents being standard normal ($z\sim \mathcal{N}(0,\mathbf{I})$) which we can use to compare to other methods of dimensionality reduction. 
However it is important to note other latent structures are possible to model, including mixing categorical and other distributions as relevant for the visualization technique and kind of analysis being done. 

The task for the encoder then is to take the nested structure $M$ as input, and output the reparamtetrized latent distribution parameters $\mu_z$ and $\sigma_z$ for each $z_i$ at the maximum resolution latent scale. In order to do this, we must choose a network architecture that can adequately handle the structure of $M$ and/or perform some transformations on $M$ to ensure it is compatible with the chosen encoder network structure.  
The approach taken by \NestedFusion is to convert the hierarchical set of heterogeneous data points into a single sequence of tokens that can be used as input to an encoder model.\footnote{The ordered sequence is an effective encoding for the nested structure as it can maintain locality within each scale and sequence models in language are perhaps the best examples of contemporary models which effectively encode nested structures (grammar in the case of language).} This is done by first using a learned mapping $T_X$ which is a linear transformation for each data scale to a common high dimensional token dimension (determined as the sum of all dimensionalities of each data modality to ensure no bottleneck at this stage) that can be used as a common shape for the encoder sequence. Then a sequence is built where starting at the lowest resolution dataset $X$ in $M$, for each data point $x\in X$ we append the corresponding token at the front of the sequence, and then find the sequence for the nested scales of that token recursively and append the resultant nested sequence (here using addition/summation notation to represent sequence concatenation).
\begin{align}
    Seq(x^{\varnothing}_{i}) &:= [T_X(x^{\varnothing}_{i})]\\
    Seq(x_j) &:= [T_X(x_j)] + \sum_{x'\in \eta(x)} Seq(x')
\end{align}

Once a sequence of tokens is generated this sequence is passed into some sequence-to-sequence encoder model which outputs a sequence of corresponding estimated latent parameterization means and variances. However only the output positions actually corresponding to $x^{\varnothing}_{i}$ inputs are then taken to sample a latent from the reparamtetrized distribution $z_i \sim \mathcal{N}(\mu_z,\sigma_z)$.

For decoding, remember the conditional distribution for data points defined as $p(x|\beta(x))$. 
Thus, what is required for decoding is a unique model for each scale in $M$, where a model either takes as input a single latent in the case of the maximum resolution scale or a set of latents as defined by the correspondence set $\beta$. 
For the latent scale decoder, a simple multi-layer perceptron is an appropriate architecture, while for the higher levels needing to decode sets of latents we can use transformers \cite{vaswani2017attention}. 
Importantly, in order to the prevent the potential pitfall of the model merely using positional information to encode information only used in the aggregate decoding step not corresponding to the actual specific latent at each point, 
our approach uses a transformer without positional embeddings in this step as they are order invariant, thus ensuring that the full distribution of latents, rather than a few arbitrary picked out latents, properly encodes lower resolution aggregate information.

Finally, given the encoder and decoder models, as well as the latent prior distributions, the models are trained using stochastic variational inference on the evidence lower bound as is standard for a VAE based architecture\cite{kingma2013autoencoding}; implemented in our case using the probabilistic programming framework, Pyro\cite{bingham2018pyro}.

To evaluate our method of Nested Fusion we test the model performance on the real, large-scale Mars Perseverance PIXL dataset introduced in Section \ref{sec:dataset-overview} comparing to existing dimensionality reduction and latent analysis techniques. As analysis of this unique dataset representing the frontier of Mars exploration is the raison d'être for this work as a whole, we specifically focus on evaluation with direct relevance towards the scientific goals and capabilities of scientists actively working at NASA JPL and around the globe on this data.

First, in order to utilize nested fusion we have to define the relevant \multidataset formulation for the PIXL dataset, which we define as:
\begin{align}
    M_{PIXL} &:= (X_q,((X_p,\varnothing),\eta_{qp}))
\end{align}  
This includes $X_q$ which consists of 103,005 of quantified spectra which are represented as 52 dimensional non-negative real valued vectors whose elements are the elemental weight percentage values produced from PIXL XRF scan points. Here $X_p$ is the set of 1,983,506 MCC multispectral imaging pixels which are 16 dimensional non-negative real valued vectors\footnote{This number is less than what you would expect given that each scan point with a quantification covers an area of 100 pixels, however in reality many of these areas overlap, meaning the same pixel can be included in multiple different scan points. Our formalization of \multidataset{s} allows this without issue and in fact it is preferred to strict partitioning as we can better models the actual resolution of dependency for each measurement. The only issue occurs when converting back into physical space such as with the color plot from Figure \ref{fig:teaser-figure}. We address this by simply averaging the multiple produced pixel level decoded inferences for overlapping pixels, however introducing more sophisticated techniques of dis-aggregation is a very promising direction for future work}.
Finally we have $\eta_{qp}$ which is the nesting function of XRF scan points to corresponding pixels. This is generated by utilizing the known range of XRF beam diameters of the PIXL instrument being approximately 150 microns, as well as the calibrated location alignment of MCC images with XRF scan points. This alignment allows us to have a shared coordinate system and thus calculate physical distance between scan-point centroids and MCC pixels. Thus we can define the nesting function to select all pixels within 75 microns of an XRF scan point, which results in the 100 pixel aggregations previously discussed: 
\begin{align}
    \eta_{qp}(x_q) = \{x_p\in X_p \;|\; \text{distance}(\text{loc}(x_q),\text{loc}(x_p)) \le 150\mu m\}
\end{align}


\subsection{Comparing with Alternative Models}
\label{sec:alternatives}
To demonstrate the effectiveness of Nested Fusion, we compare it with alternative dimensionality reduction models that can combine both scales of data. Since this problem is non-standard we must introduce the set of alternative models that allows utilization of existing methods to our problem.
We categorize these models into three types based on how they handle the nested structure of the PIXL \multidataset, Nested Fusion (our method), Concatenative Models, and Joint Models. 
We describe these three classes of models in Figure \ref{fig:pixl-graphical-models} using the language of Bayesian graphical models, which illustrates how these classes encompass a full taxonomy of problem conceptualizations for \multidataset{s}\footnote{This set only covers all possible alternatives when limited to two nested scales such as PIXL, when increasing the number of nestings the number of alternative methods produced by combining Joint and Concatenative models become combinatoric}. However within each of these classes any particular model type (e.g., UMAP or VAE) can be used. 
For our comparisons we took both alternative modeling frameworks and for each trained three representative models. First representing the most common approach to dimensionality reduction used ubiquitously in practise is Principle Component Analysis (PCA). Then to represent state of the art dimensionality reduction we used UMAP\cite{mcinnes2020umap} over t-SNE\cite{van2008visualizing} as it provides state of the art performance, has a well documented history of applications in science, and is among the techniques least sensitive to hyperparameters, and is much more computationally efficient for our scale of data. Finally we also trained a variational autoencoder to represent the most standard approach to generative latent analysis. 
Since \NestedFusion and the variational autoencoder methods are agnostic to the specific neural network sizes and architectures used, for our evaluation we trained multiple networks using simple multi-layer perceptron models (with the exception of using a transformer encoder for the \NestedFusion decoding step as described in Section \ref{sec:method}) with hidden layer sizes from 64 to 256 and a number of hidden layers from 4 to 16 and selected the best-performing models at each latent dimensionality. 
\NestedFusion{}'s open-source repository provides the pretrained tested models at \url{https://github.com/pixlise/NestedFusion}.
These methods together cover the most common latent analysis and dimensionality reduction techniques used in practice, including both parametric and non-parametric methods. 
Furthermore,
as PIXL scientists are the ultimate users who visualize the latents in 1-, 2-, and 3- dimensions we compare \NestedFusion{} with these alternatives at such dimensions.

\paragraph{Joint Models}
The first class of alternative model we will consider are joint models which attempt to model the joint distribution of a low resolution data point and it's entire corresponding nested scales in a single latent. 
For the PIXL dataset we can describe this framework as trying to find a single latent for each XRF scan point:
\begin{align}
    \forall x^q_i\in X_q. \;\; & z_i \leftrightarrow(x_i^q,\eta_{qp}(x_i^q)) 
\end{align}

\paragraph{Concatenative Models}
The other class of model considered are concatenative models, where each high resolution data point is used as the latent scale, and lower resolution corresponding measurements are simply concatenated to the high resolution sample vector. 
For PIXL we describe this as taking each XRF scan quantification and duplicating it and concatenating on top of each individual MCC pixel and using this to learn a high resolution latent:
\begin{align}
    \forall x^q_i\in X_q.\; \; \forall x^p_j\in \eta_{qp}(x^q_i).\;\; & z_j \leftrightarrow(x_i^q,x^p_j) 
\end{align}

\begin{figure*}
    \centering
    \includegraphics[width=0.85\textwidth]{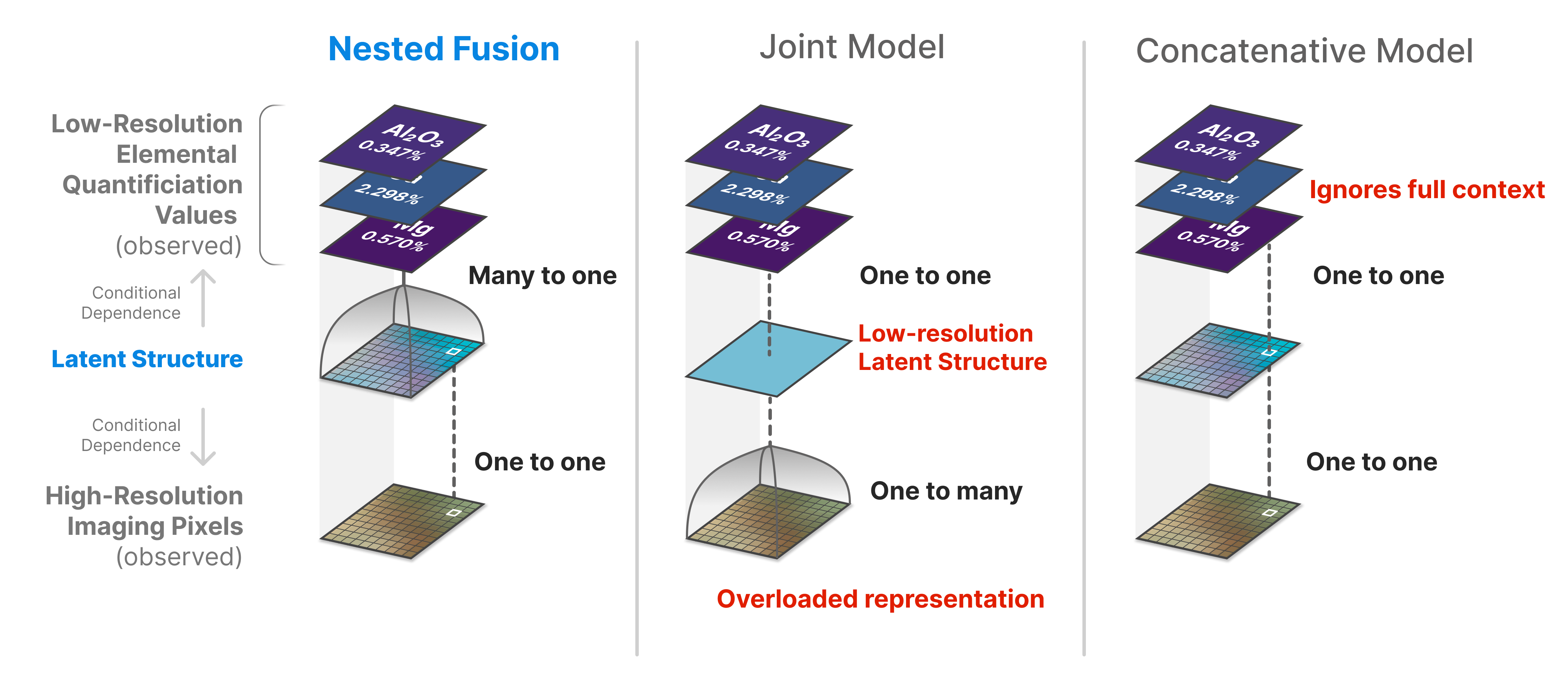}
    \caption{Comparison between alternate models and their relative downsides. The left column shows the dependence mappings from the learned latent spaces to the two measurement spaces for Nested Fusion. The center column shows how a joint encoding learns a lower resolution representation which overloads the decoder for high resolution imaging data. The right column shows how a concatenative model ignores to full spatial context of the low resolution measurements by only forming a mapping from a single high resolution point. }
    \label{fig:alternate-model-illustration}
\end{figure*}

\section{Evaluation:\\ Nested Fusion Effectiveness}
\label{sec:evaluation}

\subsection{Conceptual Drawbacks of Alternative Methods Compared to \NestedFusion{}}
\label{sec:drawbacks}
Despite covering the full set of possible alternative approaches (given the \multidataset framework), each of these method classes has substantial conceptual drawbacks, illustrated in Figure \ref{fig:alternate-model-illustration}. 
A joint model has a much more difficult encoding task where each latent value is overloaded with encoding the whole set of $\eta(x^q_i)$ making fidelity with low latent dimensionality very difficult.  
Furthermore it will also only produce a latent at the lowest possible resolution, the exact opposite of the high resolution latents in \NestedFusion. 
Concatentative models can perform somewhat better, as they produce latents as similarly high resolution to \NestedFusion. 
However the concatenative method of combining layers erases all scale contextualization of each high resolution data point, thus encoders and decoders do not have access to more complex distributional information within each nesting scale, which potentially can have an effect on the accuracy of final low resolution estimates when such information is important. 
For instance, if we consider a case where two scan points includes the same kinds of minerals but in different proportion, this will affect the values of $x_i^q$ and $x_j^q$ in such a way that any concatenative model must necessarily produce different embeddings even for the exact same kind of mineral! This false encoding of the confounding distributional information on the individual scale is inextricable from the concatenative method. However since  \NestedFusion has access to this distributional information for its encoder and decoder, in principle it could learn something close to a 'true embedding' which the concatenative model strictly could not. 
Therefore, given these conceptual drawbacks of the entire range of alternate models to \NestedFusion, we have reason to prefer it based on our prior and theoretical understanding of what the different techniques can learn in principle.

\subsection{Qualitative Evaluation}
\begin{figure}
    \centering
    \includegraphics[width=0.95\columnwidth]{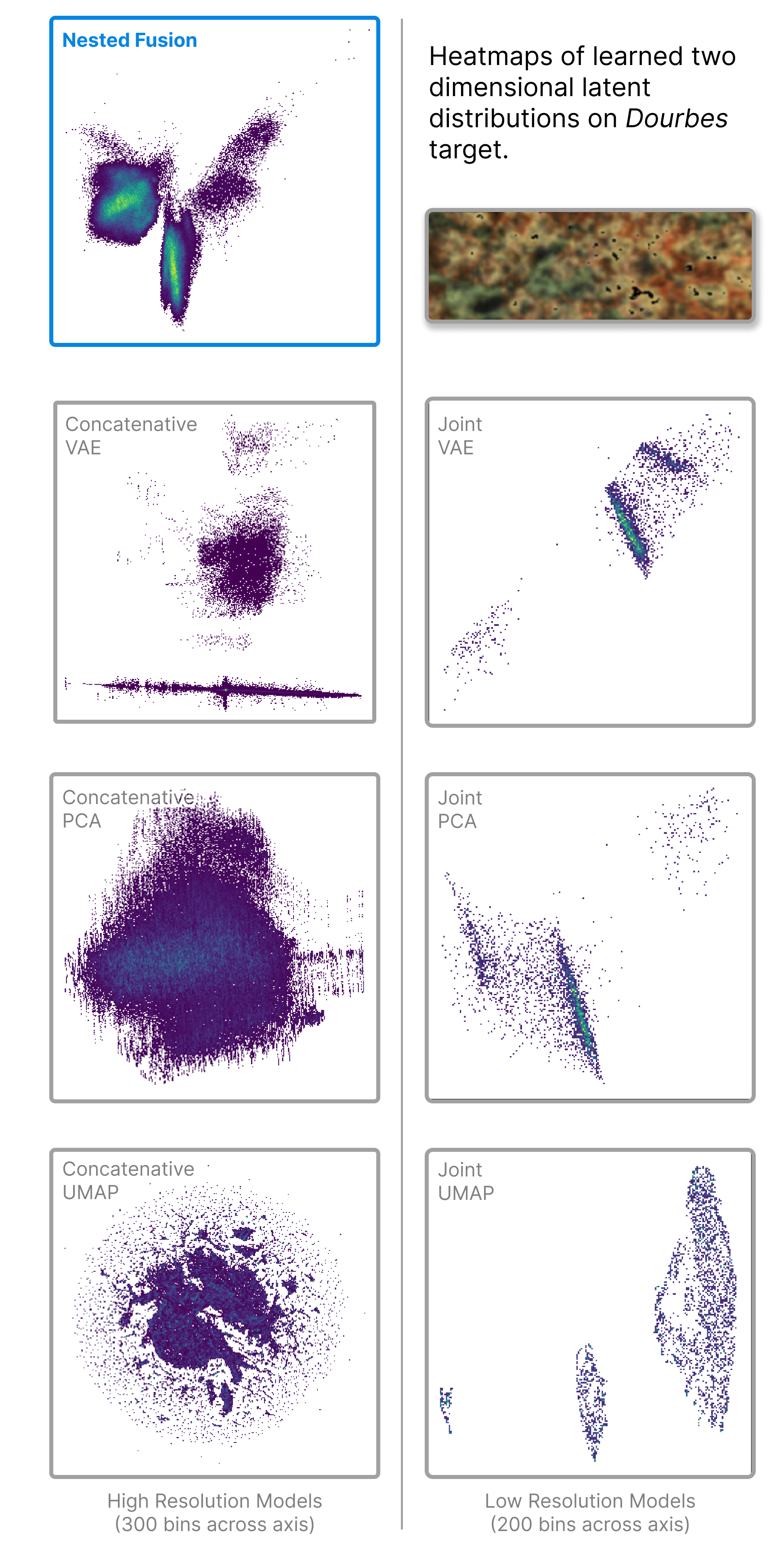}
    \caption{Comparison of 2D Latent Distributions from different methods applied to \textit{Dourbes} target (RGB map of MCC Image shown in top right). Axes are unitless latent values. High resolution models (left column: Nested Fusion and concatenative models ) displayed with 300 bins across each axis, while low resolution joint models (right column) has 200 bins due to the differing number of samples in each model type. 
    }
    \label{fig:scatter-comparison}
\end{figure}


It is important to restate that the success or failure of any of the presented latent analysis and dimensionality reduction techniques is determined entirely within the context of their actual use, for the purposes of this paper being in their application within PIXL science. 
Previous work has outlined the basic structure of how machine learning techniques have been successfully applied within the PIXL science team, by enabling an iterative semantic phenomena modeling process\cite{wright2023lessons} that helps scientists map out the space of considerations before continuing with standard domain modeling. 
Therefore we begin our evaluation of the different methods of latent analysis at the same point that PIXL scientists begin their analysis by visualizing the resultant latent distributions produced by each method directly, as two dimensional heatmaps, in order to try to discover the distinct phenomena to consider in their later modeling.
Figure \ref{fig:scatter-comparison} shows the output of each of the methods applied to the \textit{Dourbes} target from Figure \ref{fig:teaser-figure}. 
%
Specifically, for such a two dimensional heatmap plot of the latents scientists expect to see a small number of distinguishable regularities which can either be regions visualized in the heatmap as distinct areas of higher density in bright green or as separable clusters which need not be high density but otherwise must be otherwise identifiable as a standalone feature to consider. 



In Figure \ref{fig:scatter-comparison}, notice how all of the joint methods (right column) learn a comparatively small set of regularities, each showing only three distinct modes. While this regularity and differentiability is certainly a positive, we know from previous authoritative analysis on this specific target \cite{tice2022alteration} that there are at least more than three relevant phenomena that must be distinguished and so we have reasonably high confidence that these representations are overly abstracting.
The high resolution methods (left column) are more varied. 
Concatenative VAE, like the joint models, produces three primary clusters, while
Concatenative PCA encodes a continuous global structure with limited local differentiation, which in this context makes mineral identification much more difficult. 
Finally concatenative UMAP produces an extremely complex distribution which shows no consistent high-density regions. Like a Rorschach inkblot, such complexity cannot serve as a reliable basis for building trustworthy shared interpretations between scientists focused on finding specific, repeatable, and understandable regularities.
Indeed, the UMAP visualization produced in this context is perhaps the least scientifically helpful of all the options for PIXL scientists working on mineral identifications.     
Finally, we see that \NestedFusion  (top left) produces the most distinguishable structure consisting of two large high-density regions on the left (which each are themselves clearly composed of a mixture of multiple overlapping but non-identical modes) accompanied by two more lower-density clusters on the right and another on the left.
The distribution produced by \NestedFusion matches the scientific priors much more closely, where a reasonable number (more than three and less than a few hundred) of identifiable regularities likely corresponding with minerals can be clearly seen. 

\begin{figure}
    \centering
    \includegraphics[width=0.95\columnwidth]{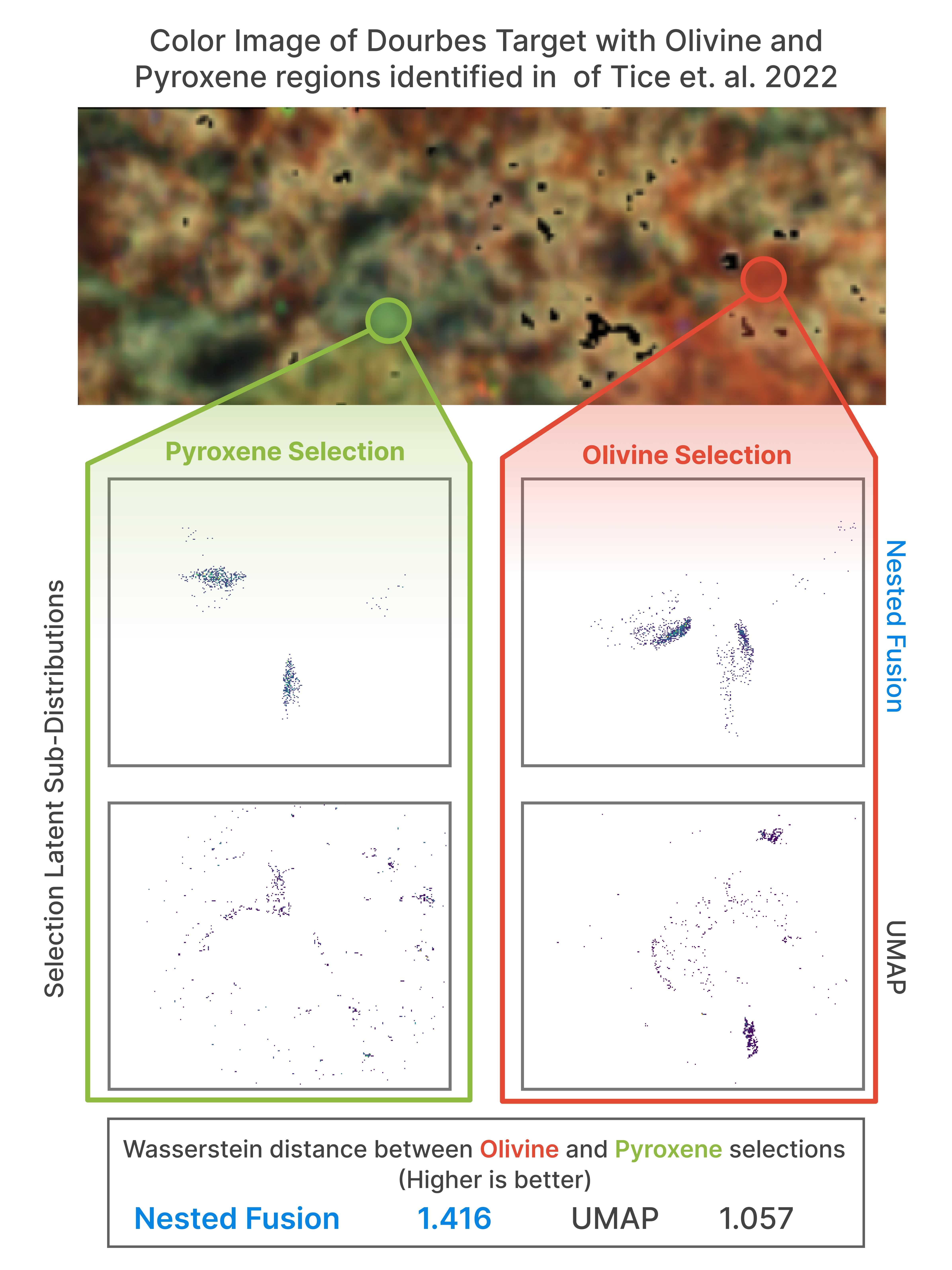}
    \caption{Comparison of Nested Fusion and Concatenative UMAP wit latent dimension 2 in differentiating distinct minerals in the Dourbes target. In green is shown a region of the target identified as Pyroxene while in red is a region identified as Olivine based on existing analysis\cite{tice2022alteration}. Comparing the latent sub-distributions of these two samples, Nested Fusion produces a distribution which has a greater degree of separation between the different minerals.}
    \label{fig:science-comparison}
\end{figure}

To further explore this effect, we compare the latent sub distributions of the highest performing methods (UMAP and Nested Fusion) when selecting known mineral grains to see the reliability of how well the latent space can be used to identify minerals.
Based on existing well analyzed data in the Dourbes target\cite{tice2022alteration} we looked to compare methods based on how well they could differentiate known distinct minerals.  In Figure \ref{fig:science-comparison} the red region corresponds to known olivine while the green region corresponds to known pyroxene, two highly distinct mineral types present at the Dourbes target. We then can select the sets of latent samples corresponding to these two spatial regions in the dataset and compare each latent sub-distributions. What we want to see is a high degree of differentiability between the distributions of these two classes of mineral. Using the Wasserstein Distance metric\cite{ramdas2017wasserstein} for empirical distributions, we found the Nested Fusion distance to be 1.416 while UMAP was 1.057. This shows Nested Fusion performing nearly \textbf{50\% better} than UMAP on this metric of mineral differentiability. Owing to the diffuse structure of the UMAP embedding when compared to the highly dense structure of Nested Fusion, the space was less clearly able to form distinct modes of different minerals, which is the primary goal of utilizing dimensionality reduction in this application context. 

These results show how Nested Fusion produces a distribution more effective at  identifying and representing distinct minerals or other phenomena which aligns precisely with what PIXL scientists hope to achieve in the scientific workflow of exploratory analysis, showing qualitatively the clear superiority of \NestedFusion to the alternative methods in assisting effective science. 

\subsection{Quantitative Evaluation}
Besides the qualitative properties of the distributions that make them practically scientifically useful, PIXL scientists also require that the latent models are trustworthy enough in retaining most of the meaningful information present in the underlying data, and since we do not know a-priori what is or is not meaningful we must ensure that a representation retains as much information as possible about the original data to reconstruct it completely. Good fidelity then is a \textit{necessary} but \textit{not sufficient} condition for effective utilization, in particular considering the fidelity of quantifications which scientists trust as more authoritative when grounding mineral identification.
Thus, we compare \NestedFusion{} with alternative models using reconstruction fidelity, a standard metric in evaluating auto-encoding models, 
to quantify how much information is preserved in the latent encodings. 
For each model we calculate the coefficient of determination $R^2$ for both $\hat{X_q}$ as well as $\hat{X_p}$ reconstructions in Table~\ref{tab:results}. 

 Our results show that Nested Fusion significantly outperforms all joint models (Joint VAE, Joint UMAP, and Joint PCA) at each reduced latent dimensionality used by PIXL scientists. 
This is expected, as explained Section~\ref{sec:drawbacks}, because the same dimensional latent values are tasked with a much greater amount of encoding and thus would be expected to perform worse at the low dimensionalities tested, and it confirms the observations from the qualitative evaluation that important information is likely being lost in the encoding. 
Concatenative models however tend to perform relatively better in these metrics. 
Among the concatenative models, concatenative PCA performs universally worst across all metrics, which is not surprising due to PCA being a linear model with limited modeling capacity. 
Concatenative VAE and UMAP both perform similarly in reconstructing the imaging layer as effectively as \NestedFusion. 
However, this layer contributes significantly less towards building trust for scientific interpretations as a standalone measurement but is most effective only when augmented with the more solid source of scientific semantic grounding in the XRF quantifications. 
When considering then the quantification reconstructions, what we find is that as predicted in Section \ref{sec:drawbacks} \NestedFusion significantly outperforms concatenative VAE in reconstructing the XRF quantification layer. Finally, concantenative UMAP's $\hat{X_q}$ reconstruction fidelity is lower but comparable to \NestedFusion{}'s --- however, given the other significant drawback of UMAP's inability to use this accuracy to practically assist in scientific exploration, its reconstruction performance is essentially irrelevant.

In summary, \NestedFusion attains higher reconstruction fidelity than the state of the art in dimensionality reduction and latent modeling while producing substantially more useful latent codes for scientific analysis. 

\begin{table}
    \centering
    \begin{tabular}{l|cc|cc|cc}
    \toprule
          & \multicolumn{2}{c|}{$\text{dim}(z)=1$} &  \multicolumn{2}{c|}{$\text{dim}(z)=2$} & \multicolumn{2}{c}{$\text{dim}(z)=3$}  \\
          Model & $R^2_p$ & $R^2_q$ & $R^2_p$ & $R^2_q$ & $R^2_p$ & $R^2_q$ \\\midrule
         
\textbf{Nested Fusion}&  0.88& \textbf{0.92}& \textit{0.97}& \textbf{0.98} & 0.97& \textbf{0.99}\\

\hdashline

Joint VAE&  0.63&0.59&0.81&0.86& 0.84&0.93\\
Joint PCA& 0.74&0.02& 0.75&0.44& 0.75&0.64\\
Joint UMAP&-&-& 0.68&0.65& 0.70&0.63\\

\hdashline
Concatenative VAE& \textit{0.89}&0.81&0.94&0.90& \textit{0.99}&0.93\\
Concatenative PCA& 0.87&0.02& 0.88&0.47&0.89 &0.65\\
Concatenative UMAP&-&-& 0.96&0.96&0.98&0.97\\

\bottomrule

    \end{tabular}
    \vspace{0.5em}
    \caption{Model reconstruction fidelity, measured as reconstruction fidelity $R^2$ values for both the MCC imaging layer $X_p$ (denoted as $R^2_p$) and the XRF quantification layer $X_q$ (deonted as $R^2_q$) for latent dimensions of 1,2, and 3 needed by 
    PIXL scientists. 
    \NestedFusion{} outperforms all models across all latent dimensions on $R^2_q$ (highlighted in bold font), the crucial metric used by PIXL scientists when assessing the scientific trustworthiness of methods.
    \vspace{-3em}
    }
    \label{tab:results}
\end{table}


\section{Scientific Deployment and Impact
} 
\label{sec:impact}

The ultimate importance of \NestedFusion is not found in its evaluation metrics but in its ability to have scientific impact by assisting PIXL scientists in visualizing and exploring combinations of datasets they simply could not easily or efficiently do otherwise. 
Towards this end, we deployed \NestedFusion in multiple capacities within the PIXL science team. 
The primary method thus far scientists have been able to utilize \NestedFusion is through its standalone implementation which is now open source at {\url{https://github.com/pixlise/NestedFusion}}. 
This implementation directly works on existing and continuously incoming PIXL data, and pre-trained models are also available. 
Pre-trained models include multiple different latent dimensionalities as well as models that include latent categorical class assignments that have the latent prior distribution being a sample of some latent class from a Dirichlet prior as well as a regular continuous latent code vector which allows the model to differentiate automatically seemingly categorically distinct regions. 

With this implementation, PIXL scientists are able to easily visualize the distribution of multiple kinds of latent encodings across many targets at once. 
PIXL scientists choose to visualize these distributions in a number of ways, including direct distributions in latent space (see Figure \ref{fig:scatter-comparison}) as well as visualizing various mappings into color overlaying the target image such as the plot in Figure \ref{fig:teaser-figure}. 
These two methods together allow scientists to see both abstract as well as spatial patterns and regularities in the data.
These visualization techniques help PIXL scientists firstly discover quick heuristic understandings of the distribution of empirical phenomena present in a single target as well as commonalities in phenomena across multiple targets. 

Through participatory design sessions over 6 months with nearly a dozen scientists, we have discovered a primary 
(though not exclusive)
workflow that \NestedFusion enables within the context of exploratory data analysis. 
When a new dataset is generated, the process by which PIXL scientists begin to come to a consensus on its mineral composition is highly iterative, involving bringing forward various hypotheses and then coming up with ways to test these hypotheses given the data. 
The mechanics of how a hypothesis is tested can be complex and difficult, and so a method that could assist in having a more informed starting position in this iteration can greatly increase the efficiency of the whole process, saving a huge amount of extremely valuable and limited time. 
By forming a latent space over the whole history of PIXL data, and an encoder that can efficiently process this new data before having to retrain, new data can be quickly visualized and broken down into a few key regularities which can be compared to historical precedent of regions or even individual grains which bare a strong resemblance to regions or grains in the new dataset. 
This then helps to form a better initial assessment of the minerals present at a target and thus substantially speed up the overall identification process. This transforms the workflow of initial exploratory analysis, which historically would take the roughly 10-person team of spectroscopists approximately 21 days in collaboration to come to an initial determination of minerals into one which a single scientist can generate instantly a latent distribution and through refinement generate an identification of comparable quality in a matter of hours. 

We found how the combination of the high fidelity of \NestedFusion along with its computational efficiency at inference time were both essential components compared to existing or alternative models to achieve buy-in by scientists. Furthermore we found that non-parametric alternative methods such as UMAP proved to be ineffective despite competitive fidelity due to the inability to form distributions for new data efficiently and producing distributions that are difficult or impossible to reliably interpret within the context of looking to understand specific phenomena, and thus does not help solve the scientific workflow problem that \NestedFusion addresses.

 \NestedFusion provides a fundamentally new way for PIXL scientists to quickly visualize distributions of phenomena that span multiple measurement types and scales and thus explore new data more efficiently and effectively than was previously possible. This has provided a lesson for any interested applied data scientist: increasing the alignment between machine learning problem statement and scientific problem ontology, in this case by more accurately modeling multiple scale relationships, is an absolutely essential component of achieving genuine impact with these tools. Therefore we hope that future work will continue to develop ways in which we can improve the very frame from which we pose data science problems just as much as improving the methods for how we solve them, in order to make sure we can not only do better data science, but just do great science.

\begin{acks}
This research was carried out in part at the Jet Propulsion Laboratory, California Institute of Technology, under a contract with the National Aeronautics and Space Administration (80NM0018D0004).
\end{acks}

\bibliographystyle{ACM-Reference-Format}
\balance

\bibliography{ref}

\end{document}